# The Ancient Astronomy of Easter Island: Canopus and the Legendary King *Hotu-Matua*

Sergei Rjabchikov[1]

[1]The Sergei Rjabchikov Foundation - Research Centre for Studies of Ancient Civilisations and Cultures, Krasnodar, Russia, e-mail: srjabchikov@hotmail.com

## Abstract

Watchings of Canopus as a herald of the winter were important duties of ancient priests-astronomers on Easter Island. All the analysed data witness that this star was observed during the first and second voyages from Mangareva to the island. The names of king *Hotu-Matua* (*Anua-Motua*) and his father *Tara tahi* have been decoded. Several *rongorongo* records from the Esteban Atan manuscript have been deciphered advantageously. The new view at a painted barkcloth figurine connected with the bird-man cult has been offered. Some data about watchings of Aldebaran, the Pleiades, the sun, the moon, Venus and Mars have been collected as well.

**Keywords**: archaeoastronomy, writing, folklore, rock art, Rapanui, Rapa Nui, Easter Island, Polynesia

## Introduction

The civilisation of Easter Island is famous due to their numerous ceremonial platforms oriented on the sun (Mulloy 1961, 1973, 1975; Liller 1991). One can therefore presume that some folklore sources as well as *rongorongo* inscriptions retained documents of ancient priest-astronomers.

### The Information on the Mataveri Calendar as the Basis of the Bird-Man Cult

On a boulder at Mataveri (a crucial area of bird-man rites) some lines were incised; most of them were the directions of the setting sun according to Liller (1989). I have calculated the corresponding days for the year 1775 A.D. (Rjabchikov 2014a: 5, table 2; 2015: 2, table 1; 2016a: 1, table 1; 2016b: 1, table 1), see table 1. Here and everywhere else, I use the computer program RedShift Multimedia Astronomy (Maris Multimedia, San Rafael, USA) to look at the heavens above Easter Island.

**Table 1.** The Mataveri Calendar

June 22 (the azimuth of the sun = 296.2°): one day after the winter solstice;
July 21 (292.5°): the star Capella (α Aurigae) before dawn;
August 11 (286.7°): the star Pollux (β Geminorum) before dawn;
September 2 or 3 (277.9°): the star β Centauri [*Nga Vaka*] before dawn;
September 21 (270.1°): the day before the vernal equinox (the key moment of the bird-man feast);
September 24 (268.7°): the new moon;
September 27 (267.4°); the fourth night: the measure of the visible dimensions of the moon; the waxing crescent was well seen in the sky; one night before the beginning of the first *Kokore* lunar series;
October 1 (265.9°); the eighth night: the measure of the visible dimensions of the moon; the first quarter of the moon;
October 3 (264.7°); the gibbous moon (the 10th night); the end of the first *Kokore* lunar series;
October 22 (256.8°): near the new moon;
November 8 (250.7°): the star Spica (α Virginis) before dawn;
November 12 (249.3°); Venus as the Morning Star before dawn;
November 14 (248.7°); one night before the last quarter of the moon;
November 23 (246.3°): the new moon;
December 20 (the azimuth of Aldebaran = 339.1°): the star Aldebaran (α Tauri) at night;
December 21 (the azimuth of Aldebaran = 322.1°; the azimuth of Canopus = 177.5°): the stars Aldebaran (α Tauri) and Canopus (α Carinae) on the same night (Rjabchikov 2013: 7); the day of the summer solstice; one night before the new moon.

Canopus is the second brightest star after Sirius (α Canis Majoris). The ceremonial platform Ahu o Pipiri located on the southern coast of the island was named after Canopus. This spot has number 202 in the catalogue of Englert (1974: 271). On the map by Métraux (1940: 8, figure 1) the place is called *Opipiri*, in the other words, *O Pipiri*. The next platform (number 203 per Englert) is called *Ahu Romo*. Both *ahu* are situated near the religious complex Ahu Akahanga – Ahu Urauranga te Mahina (numbers 204, 207 per Englert). In accordance with Popova (2015), *Romi Renga* was a poetic metaphor for the star Aldebaran. The term *romo* is the variant of the term *romi* because of the gradations of the sounds *o/i* in the Rapanui language.

On the basis of the deciphered data of the Mataveri calendar it is apparent that the ancient Rapanui astronomers fixed the attention of both Canopus and Aldebaran. I think therefore that the platform Ahu Romo was named after Aldebaran. Both spots were undoubtedly small observatories to watch those stars.

### The Star *Pipiri* in the Hieroglyphic Records

According to Barthel (1958: 203), the stellar glyphs **7-7** read *Pipiri te hetuu*. Their archaic reading was *Pipiri tuu* or *Piri tuu*.

The folklore text "*Apai*" (Thomson 1891: 517-518) contains the following phrases that were investigated by Barthel:

*mato kapipiri te hetun tan aranga*;
*mata mata ka pipiri te hetu tau avanga* (the misprints in the words *hetun* = *hetuu*, *tan* = *tau* and *aranga* = *avanga* are seen well).

I have reconstructed the original sentences which have one and the same meaning:

*Mata; ka Pipiri te hetuu tau avanga*; The face (the solar symbolism); *Pipiri te hetuu* (Canopus) burnt (*ka*) from the season (*tau*) of the grave (*avanga*) (= the darkness; winter; coldness etc.).
*Matamata; ka Pipiri te hetuu tau avanga*. Ditto.

It is the description of the first appearance of Canopus before dawn in the winter.

The text "*Apai*" also contains the following sentence: *Ko hao ko Piri e Atua.* '(The star) *Piri e Atua* (*Pipiri* = Canopus) rose' (Rjabchikov 2011). So, the complete name of Canopus was *Piri* (or *Pipiri*) *tuu* [= *te hetuu*] *e Atua*. As a parallel one can mention Maori and Tahitian *Atutahi* (Canopus) < *Atua tahi* 'The first god' or *Atu tahi* 'the first mark' (the symbolism of the beginning of the New Year in June).

The text on the Tahua tablet (A) where the star-like signs **7-7** (*Pipiri te hetuu*) are repeated six times is revealed in figure 1.

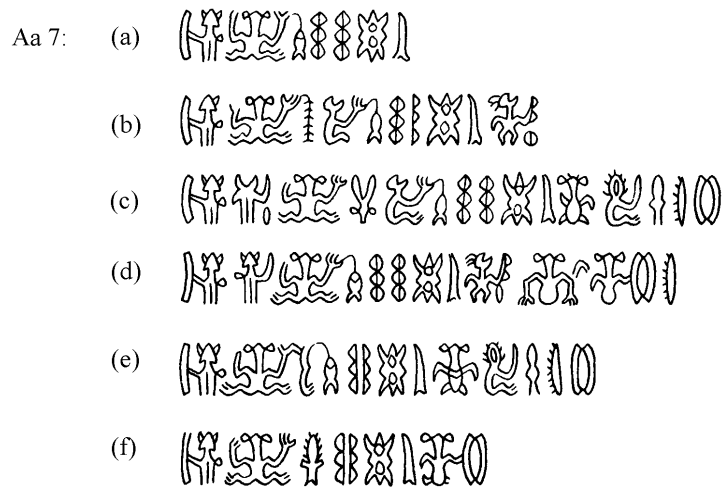

Figure 1.

Glyph **5** reads *atu* and *atua* (and *ati*, *atia*, the syllables: *tu* and *ti*). Hence, signs **7-7 5** of the full name of Canopus read *Tuutuu* [*Tuu Pipiri* or *Piri*] *Atua* (as said above), see figure 2.

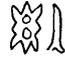

Figure 2.

One can attempt to extract the additional information from the phrase *ka Pipiri te hetuu*; in figure 3 the expressions **17-17 7-7 5** *teatea Tuutuu* [*Tuu Pipiri* or *Piri*] *Atua* are shown as they have been found on the Tahua board.

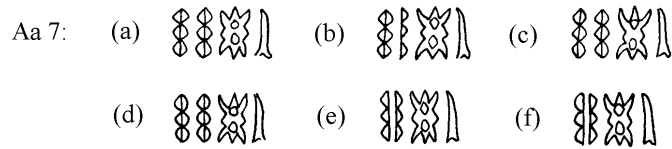

Figure 3.

The expression **17-17 7-7 5** *teatea Tuutuu* [*Tuu Pipiri* or *Piri*] *Atua* means 'Canopus appeared, shone,' cf. Rapanui *ka* 'to burn,' and *tea* 'to appear (about stars).' Here one can see different allographic variations of glyph **17** (cf. Fedorova 1975: 279, table II, the figures on the right).

The Tahua record (see figure 1) incorporates several key sections, see figure 4.

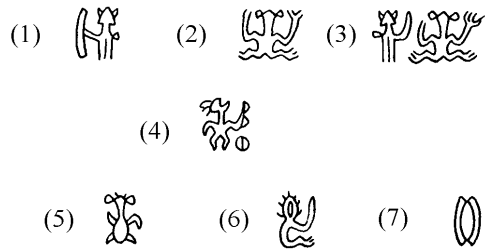

Figure 4.

(1) Glyphs **4-26/21** = **4-26 21** *Timo ako* 'The pupil is writing' (the same words as at the beginning of the famous folklore text "*He timo te akoako*"). The repeated phrases and sentences were carved by students in *rongorongo* schools. The famous chant began with the words: *He timo te akoako* (a pupil is carving [glyphs]), *he akoako tena* ([he] is learning them) (Fedorova 1978: 314-315, 361) Here Old Rapanui *timo* signifies 'to strike with a pointed instrument,' and *ako*, *akoako* 'to learn,' cf. Maori *timo* 'to strike with a pointed instrument,' Hawaiian *kimo* [*timo*] 'to strike (with a stone etc.),' Marquesan *timo* 'sign,' Rapanui *akoako* 'to learn,' Samoan *a'o*, *a'oa'o* 'to learn; to teach,' Hawaiian *a'o* [*ako*] 'to learn; learning,' Tuamotuan *ako* 'to learn,' Mangarevan *ako* 'to try, to exercise, to practise' and Marquesan *ako* 'to try, to exercise.' The Rapanui terms *akoako* 'to repeat a song over and over again to learn it,' *ako* 'to sing, to recite,' *he ako i te kaikai* 'to recite the text associated with a string figure,' and *akoako* 'to recite the hymn dedicated to a deity' had the secondary meanings from Old Rapanui *ako*, *akoako* 'to learn.' Their primary meanings were Old Rapanui *akoako* 'to learn (a text, a song) repeating it over and over again,' *ako* 'to sing, to recite (to learn the text written on a tablet),' *he ako i te kaikai* 'to learn the text associated with a string figure,' and *akoako* 'to learn the hymn dedicated to a deity.' In the same way, Rapanui *hakaatu* (*haka-* is the causative prefix here) with the secondary meaning 'to repeat' came from Old Rapanui *ati*, *aati*, *hati*, *hahati* 'to strike; to break; to write,' cf. also Rapanui *hatu*, *hahatu* 'to fold' (Rjabchikov 2012a: 565).

(2) Glyph **69** *Moko* (Lizard) and (3) glyphs **26-21 4 69** *Moko atua MOKO* (the god 'Lizard'). It is the description of the first statue of the ceremonial platform Ahu Tongariki taken from lost local astronomical diaries (Rjabchikov 2010a). The word *moko* 'lizard' is written with the help of an ideogram (glyph **69** *moko*) and two quasi-syllables (glyphs **26-21** *ma-* (or *mo-*) (*o*)*ko*). The statue represented the god *Hiro* (the new moon and the solar eclipses).

(4) Glyphs **6-113 139** *Hova raa* (= *raai*) 'the sun god *Ho(v)a*.'
(5) Glyph **68** *Honu*, *Hono*, *Ono* denotes the Pleiades (M45; NGC 1432) here. Per Lee (1992: 80), in some instances the turtle can be displayed as an emblem of the Pleiades in the Rapanui rock art.
(6) Glyphs **2 14** read *Hina Haua* 'the moon goddess *Hina Haua*.'
(7) Glyph **57** *tara* 'dawn; to dawn,' cf. Maori *tara* 'rays of the sun, shafts of light, appearing before sunrise.'

Certain of glyphs **17-17** of the word *teatea* 'to shine; to appear' are rendered on the Tahua tablet, see figure 5.

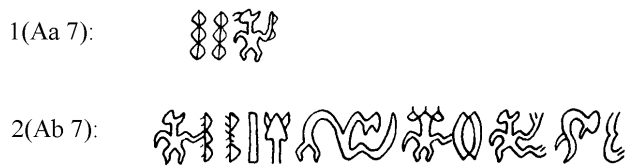

Figure 5.

1 (Aa 7): **17-17 6-113** *Teatea Hova*. The sun deity *Hova* = the sun appeared.
2 (Ab 7): **6 17-17 4 21 31 5 6 57 6 62-43** *Ha teatea atua ko Make(make)-atua, ha tara, ha tomo*. The god *Makemake* (= the sun) appeared, (the sun) peeped out, (it) entered (rose).

The sequences of glyphs **17var-17var (= 17-18, 18-18) 7-7 5** 'Canopus appeared, shone' are presented thrice on the Small Washington (R) tablet, see figure 6.

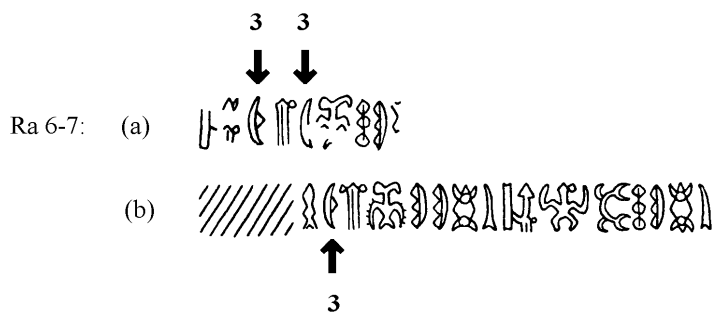

Figure 6.

Ra 6-5: (a) **4-26/21 3 26- 3 44 17-18** (a damaged segment: **7?-[7?]** …) *Timo ako: (marama!)ma(marama!)ta teatea* … (A pupil is writing: learn!) *Mata. teatea [Pipiri tuu atua]*. Canopus appeared before dawn…
(b) (a damaged segment) **73 3 26-44 18-18 7-7 5 4-26/21 6 8 17-18 7-7 5** … *(He marama:) Mata teatea Pipiri tuu atua. Timo ako: A Matua, teatea Pipiri tuu atua*. (He has learnt:) *Mata teatea Pipiri tuu atua*. Canopus appeared before dawn. (A pupil is writing:) *A Matua. Teatea Pipiri tuu atua*. (King) *Matua* [*Hotu-Matua*]. Canopus appeared.

In this record one can recognise glyphs **17** and **18** as allographic variants. The inscriptions that contain the words *timo ako* (*he timo te akoako*) are the texts of exercises in *rongorongo* schools. In such cases glyphs **3** *marama* often mean 'learn!; learnt; for the learning' (the exclamations of teachers). Thus, e.g., glyphs **3 26 3 44** read (*marama!*)*ma*(*marama!*)*ta* = *marama! mata* '(learn the word) *mata*.' Notice that a glyph between two crescent glyphs **3** in segment (a) in Barthel's (1958: Ra 6) reproduction of the tablet is unclear; I have reconstructed sign **26** *ma* there using a photo of the tablet.[2]

Glyphs **26-24** *Mata* 'Face' are the phonetic reading of glyph **60** *Mata* (cf. Rjabchikov 2016b: 12, figure 17; 13, figure 18) representing a face and in some cases denoting the sun deity *Makemake*. The read words *mata* correspond to the words *mato* (= *mata*) and *mata mata* (= *matamata*) 'the face = the sun' from the "*Apai*" text, see above. See about PMP (PPN) **mata* 'face; the sun' in Rjabchikov 2014b: 163ff.

---

[2] See that photo: <http://collections.si.edu/search/tag/tagDoc.htm?recordID=nmnhanthropology_8010183>.

# The Linguistic Background

Old Rapanui *marama* means 'school; to learn; to know' (Rjabchikov 2012a: 566, figure 2, fragments 2 and 3), cf. Rapanui *marama* 'intelligent,' *hakamarama* 'school,' Niuean *maama* 'to know; knowledge,' Penrhyn *mārama* 'wise,' Marquesan *maama* 'learnt,' and Mangarevan *marama* 'ditto.'

At the inception of one of sheets of the Esteban Atan manuscript with glyphs and brief records written by Roman letters (Heyerdahl and Ferdon 1965: figure 112) there is such a record: *HE MARAMA*. Now it is clear that it means 'The School.' It is a fail echo from pre-European *rongorongo* schools.

Let us consider some records from this late lesson book:

(1) (a) Glyphs **44-44** (Ibid., figure 111, fragment 1) *E rua tagata he oho ki roto ki te koro.* 'Two men came into a *koro* house.' The glyphs read: *Taha, taha.* 'Two (frigate) birds (= two men wearing masks of (frigate) birds).'

(b) Glyph **69** (Ibid., figure 112, fragment 5) *Tagata honui i roto i te koro.* 'Nobleman is inside a *koro* house.' The glyph reads: *Moko.* 'Member of the Hanau Momoko tribal union (Tuu 'Those who Arrived;' Miru as the new elite class, etc.).'

(c) Glyphs **44 108b-108b** (Ibid., figure 112, fragment 6) *He mo tuha o te koro: ea vaia i te kupu o te ate.* 'Piece of the chicken (*mo* = *moa*) (= the cooked chicken) for a *koro* house: the gift (*vaia* = *vaaia*) is lifted (*ea*) to record (*ate* = *ati*, *hati*) counting the time (*kupu*).' The glyphs read: *Tahirihiri.* 'Lifting.' Notice that glyph **108b** *(h)iri* is engraved together with signs of different lunar phases on a stone cylinder at Vinapu (Rjabchikov 2001: 219). Cf. the name of the ceremonial platform Ahu Tahiri.

(d) Glyphs **17 4-35** (Heyerdahl and Ferdon 1965: figure 113, fragment 9) *Ka maitaki te Huru o te koro.* 'Good (celestial) fire of the Pleiades (*Huru*, *Uru*, the name of one star of this cluster) (is seen) from a *koro* house.' The glyphs read: *Tea tupa.* 'Shine (visible) [or: Vision] from (an ancient tower) *tupa*.'

(e) Glyphs **44-68** (Ibid., figure 114, fragment 9) *Tagata iku o roto i te koro.* 'Best man inside a *koro* house.' The glyphs read: *Tahonga.* 'Priest.' In compliance with Métraux (1940: 137), Old Rapanui *tahonga* means 'expert craftsman.'

Plainly, the early *koro* houses were ancient towers *tupa*. According to Lee (2004: 36), *tupa* were exploited to watch the Pleiades. These records can shed new light upon the *koro* feasts and the functions of the *koro* houses. One can state with assurance that the cult of fertility deities and dead ancestors was the foundation of the feasts.

(2) (a) Glyph **27** (Heyerdahl and Ferdon 1965: figure 116, fragment 3) *O tapu Ara Taha.* 'Prohibition of the Way(s) of the (Frigate) Bird (for many).' The glyph reads: *Ra(h)u.* 'Fruits/Eggs figuratively.'

(b) Glyphs **44-30** (Ibid., figure 121, fragment 3) *He tagata uruuru Hoa.* 'Man of the entrance to the Friend (the statue Hoa-Hakananaia).' The glyphs read: *Tane* (the name of the sun deity, also known as *Tiki*, *Makemake*, *Rarai-a-Hoa*, *Rarai-a-Hova*).

(3) (a) Glyphs **6 16 25** (Ibid., figure 115, fragment 7) *He tagata kua hakatuu i te maro.* 'Man that comes in feathers' (Kondratov 1965: 414, table 2, the translation for source [115, 6]). The glyphs read: *A kahi (h)ua.* 'Tuna fish (the sea god *Tinirau* or gifts) of the royal staff *ua*.' Cf. Rjabchikov 2014b: 167.

(b) Glyphs **6 48 14** (Heyerdahl and Ferdon 1965: figure 116, fragment 9) *He tangata e gua i te ao.* 'Man recited: the authority.' The glyphs read: *Hau HAU.* 'King.'

(c) Glyphs **49 25** (Ibid., figure 119, fragment 5) *Kua ki te tagata mee vanavana.* 'Man said: a thing for the war.' The glyphs read: *(Ariki) mau (h)ua.* 'King with the royal staff *ua*.'

(4) (a) Glyphs **115 44** (Ibid., figure 116, fragment 8) *Kua puru te ika i roto i te hare paega.* 'Hiding the victim in the *paenga* house' (Kondratov 1965: 414, table 2). The glyphs read: *Taka Taha.* 'Solar circle of the (Frigate) Bird.'

(b) Glyphs **5** (= the inverted sign) **12** (Heyerdahl and Ferdon 1965: figure 117, fragment 3) *Paoa tia nua.* 'Guard pierced a garment.' The glyphs read: *Ti ika.* 'Murder of a victim.'

Glyphs **27 5** (Ibid., figure 118, fragment 2) *Kua ki te tagata ka ava tau umu.* 'Man said: a priest (*tau*) stroke (*ava*) (a victim to cook in) an earth oven.' The glyphs read: *Rau ti.* 'Leaves of the murder.' Old Rapanui *taura* means 'priest,' cf. Rapanui *tahutahu* 'wizard' and *taua* 'warrior' as well.

(c) Glyphs **65 3** (Ibid., figure 120, fragment 9) *Ka ea tau Pupuhi, ka oho.* The priest (*tau*) Pupuhi (*Puhipuhi*) elevated himself, (then he) went. The glyphs read: *Rangi Hina.* '(The priest) *Rangi Taki* [The Counting (= the lunar phases) in the Sky].' The name of the priest *Rangi Taki* is mentioned in a Rapanui legend (Felbermayer 1971: 29-32).

(5) (a) Glyphs **14 4** (Heyerdahl and Ferdon 1965: figure 116, fragment 2) *Pua ka ravaravato take tamahahine*. 'Egg (= *mamari* = *komari* or vulva figuratively) is added during the girl *take* (initiation) ceremony.' Besides, it is known that Rapanui *pua* 'flower' is a metaphor for 'woman' (Barthel 1978: 45). The glyphs read: *Haua atua*. 'The moon goddess *Haua*.'

(b) Glyph **105** (Heyerdahl and Ferdon 1965: figure 116, fragment 5) *He paki, poki*. 'Child [son, daughter], child [son, daughter]' (the gradation of *a/o*). The glyph reads: *Moa* 'Cock; son figuratively.'

(c) Glyph **44** (= the inverted sign) (Ibid., figure 121, fragment 1) *Poki manu*. 'Child during a *manu* (bird) ceremony (on the islet Motu Nui).' The glyph reads: *Taha*. '(Frigate) Bird.'

(d) Glyph **11** (Ibid., figure 121, fragment 2) *Poki take*. 'Child during a *take* (initiation) ceremony (at Orongo). The glyph reads: *Poki* 'Child; son; daughter.'

## The Name of King *Hotu-Matua* in the *Rongorongo*

Consider the record on the Tablette échancrée (D), see figure 7.

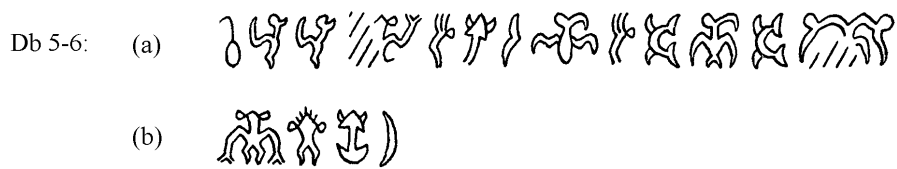

Figure 7. (The drawing is corrected.)

Db 5-6: (a) **103 19-19 69 58 21 50 44 58 8-44 8-44** *Pea, kiki, moko, taiko i ta(h)a tahi, matua, ta(h)a, matua, ta(h)a!* Write, speak, write, repeat (the glyph) *ta(h)a* at first, (then the glyph) *matua* (after the glyph) *ta(h)a*, (the glyph) *matua* (after the glyph) *ta(h)a*!
(b) **6 75 3 21 3** *Ako marama, ako marama*. The pupils were learning, the pupils were learning.

I have proposed that in this record the name of the first legendary king *Hotu-Matua* is mentioned (Rjabchikov 2012a: 565-566, figure 1).
Consider the record on the Tahua tablet, see figure 8.

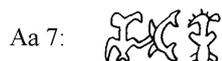

Figure 8.

Aa 7: **44-8 49** *Tava-Matua (ariki) mau*. King *Tava-Matua* (= *Hotu-Matua*).

Glyph **44** reads *taha, taa*, and *ta* as a morpheme or part of different words; in some cases this sign denotes the name of different birds: *manu* 'bird,' *taha* 'frigate bird,' *tavake* 'tropic bird,' *kena* 'booby,' *manu-tara* 'sooty tern' and so on.

Barthel (1978: 55) has investigated the names of explorers who had been sent by king *Hotu-Matua* from the distant homeland Hiva to Easter Island. Among them five ones were sons of *Hua Tava*: *Kuukuu a Hua Tava*, *Ringiringi a Hua Tava*, *Nonoma a Hua Tava*, *Uure a Hua Tava*, and *Makoi a Hua Tava*.

Butinov (1984) has read the words *hua Tava* as 'son of *Tava* (= *Tavake* 'The Tropic Bird'),' cf. Rapanui *hua atahi* [*hua a tahi*] 'only son.' Hence, the name *Kuukuu a hua Tava* means '*Kuukuu*, a son of *Tava*,' the name *Ringiringi a hua Tava* '*Ringiringi*, a son of *Tava*,' the name *Nonoma a hua Tava* '*Nonoma*, a son of *Tava*,' the name *Uure a hua Tava* '*Uure*, a son of *Tava*,' and the name *Makoi a hua Tava* '*Makoi*, a son of *Tava*.'

Moreover, the scholar came to a conclusion that that the name *Tava* and the name *Nga Tavake* from the legend "*Ko Hau Maka*" (Heyerdahl and Ferdon 1965: figures 129-130) were epithets of king *Hotu-*

*Matua*. By the bye, from the legend it is apparent that *Nga Tavake* and *Hau Maka* (King 'Stone') were relatives, either brothers or a son and his father.

According to recent investigations, the crew of king *Matua* (*Motua*) from Mangareva settled on Easter Island in 1000-1300 A.D. (Buck 1938; Martinsson-Wallin and Crockford 2001; Martinsson-Wallin 2010). I have studied this problem in my works (Rjabchikov 2010b; 2014c; 2014d). One can presume that at least two expeditions were from Mangareva: at first the chief *Taratahi* sailed to an island called *Mata-ki-te-Rangi* (Easter Island). Later on his son *Anua-Motua* together with his sons (among them was the priest *Angiangi*) swam to the same island in a double canoe. In the memory of Easter Islanders they displayed as the first expedition of the explorers who did not return to the island Hiva and the second expedition of king *Hotu-Matua* and some other important characters.

The names *Anua-Motua* and *Hotu-Matua* have the common root *motua* (*matua*) 'father' (Martinsson-Wallin and Crockford 2001). I surmise that the name *Tara tahi* means 'One *Tara*' or 'The first *Tara*.' Some features of the personage *Angiangi* (< *Angi*) could be retained in the personage *Ringiringi* (*Ri* or *Riri Angi*, cf. Rapanui *riri* 'anger,' and Tahitian *ani* 'request, petition').

In the Rapanui folklore text known well as the Creation Chant the following sentence is presented: *Matua-Anua ki ai ki roto ki a Pipiri-hai-tau, ka pu te miro* (Métraux 1940: 320-322). In my opinion, this text can be translated as follows: '*Matua-Anua* (= king *Hotu-Matua*) by copulating with the star Canopus of a (certain) season produced the canoe (ship).'

It is the basic linguistic evidence that the Mangarevan voyager *Anua-Motua* and the Rapanui voyager *Hotu-Matua* were one person. The connection between king *Hotu-Matua* and the star *Pipori* is fixed. Some *rongorongo* records about that event will be deciphered below.

**The Linguistic Background**

Here three inscriptions containing glyph **49** *(ariki) mau* are rendered. In the first two cases it denotes the king, and in the third case it is included in the word *maunga* (mountain).

Consider the record on the Mamari tablet (C), see figure 9.

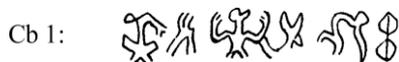

Figure 9.

Cb 1: **49 15-25 6 123 (= 27) 44b-17** *(Ariki) mau rau, ha rau tua tea*. The king multiplied the number of the sweet potatoes (*tua tea*). {The reading of glyph **49** *ma(h)ua* (to be abundant) is possible, too.}

This text correlates with the Rapanui chant about the power of the king (Métraux 1937: 52-54). The term *Tua tea* (The white back literally) was the name of a variety of sweet potatoes.

Consider the record on the Tablette échancrée, see figure 10.

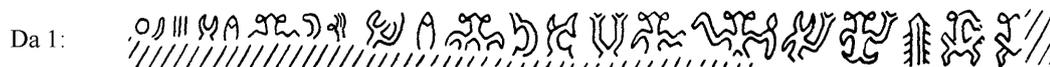

Figure 10.

Da 1: **139 4 34 59 12 6 35 58 6 12 6 8-8 53 68 35 68 30 11 69-50 26 49-49** *Teke. Tuura ka iko a pa tahi, ha iko a reirei, maro, honga pa honga ana, poki mokohi ma (ariki) mau-(ariki) mau*. It is a royal hat (*hau teketeke*). A servant (priest) *tuura* (= *taura*) took the *rongorongo* inscription (*pa* = the carvings literally) at first, (he) took pendants *rei-miro*, a long feather garland *maro*, figurines *tahonga* of the fertility, a young virgin (*poki mokohi* = *neru*) for the great king.

This archaic text was preserved in a late song (Felbermayer 1972: 276; Fedorova 1978: 356).

Consider the record on the Santiago staff (I), see figure 11.

I 3: (a) 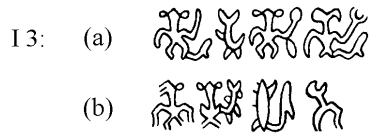

(b)

Figure 11.

I 3: (a) **6 (102 123) 6-4 44** *A Hatu Taha*. (It was the god) *Tiki-te-Hatu* (looking like) the (Frigate) Bird.
(b) **6 49-(123)-28 (102) 62** *A Maunga Toa*. (It was) Mount (Maunga) Toatoa.

It is the important narration about the cult of the solar god *Tiki-Makemake*, the main god (together with the god *Rongo*) of the eastern tribal union. The biggest glyph **1** *Tiki* was engraved on a stone near the ceremonial platform Tongariki [Tonga Ariki]. We are supplied with one incident when a priest in his prayer for the rain called the name of this great god (Felbermayer 1971: 29-32).

Per Routledge (1998: 224), six men called *Nga Ruti Mata Keva* of the eastern tribe Tupa-Hotu lived in a cave of Mount (Maunga) Toatoa on the southern shore. They were the members of a secret society, painted their faces in red, white and black, and communicated that they were gods. I have translated the words *mata keva* as 'mask; painting of a face.' The nickname *Ruti* means 'November chiefly.' Cf. two parallel versions of this tradition in Métraux 1940: 383, 139-140. In the last tale the six young men *Aamai* and their relative *Toka-ngutu* acted. All wore masks and declared that they were the god *Makemake*.

Rapanui *ngutu*, the second part of the name (nickname?) *Toka-ngutu*, means 'beak.' Old Rapanui *toka* (= *toko*) 'great god; deity' corresponds to Maori *toko* 'sacred pole or stick set up in honour of a deity.' Moreover, Maori term *toko* denotes several gods including *Tane* and *Tangaroa* (Te Rangi Hiroa 1949: 467). Thus, *Toka-ngutu* was a person wearing a bird mask of the god *Tiki-Makemake*. In the archaic times such a man could be a bird-man who obtained plentiful gifts (the tuna fish etc.). The nickname *Aamai* (*A mai* < *A mahi*) means '(Man) who got a gift,' cf. Rapanui *mahia* 'gift.'

The red colour was the notation of that deity (see the details in Rjabchikov 2000). It is my belief that the bird masks were painted in fact. The white colour (cf. Rapanui *tea* 'white') denotes day, the sun, summer, and the black colour (cf. Rapanui *uri* 'black') denotes night, eclipse, winter. Let us examine the figurine made of barkcloth, reeds and wood, and housed in the Peabody Museum, Cambridge, Massachusetts (Métraux 1940: 238, figure 30; Heyerdahl 1976: plates 19 and 20; Orliac and Orliac 1995: 21, figure; Kjellgren 2001: 59-60, figure 24). In my opinion, it is an image of a man played the role of the god *Tiki-Makemake* and worn a bird mask. This mask – the face of the sacramental puppet – was painted orange (= red/yellow), white, and black. On the chin of the figurine we can see glyph **34** *raa* (the sun). On the neck of the figurine there are signs depicting the roads (cf. Rjabchikov 2012a: 567, figure 4; 568, figure 5). On the breast of the figurine we can see glyphs **84 39** *ivi raa* 'the ancestor – the sun.' The symbols of different ways correlate with the phrase *Ara Taha* (The Paths of the (Frigate) Bird) recited in the end on the famous chant "*He timo te akoako*" (Rjabchikov 2016b: 15).

## The Addition Data about King *Hotu-Matua*

Consider the record on the Tahua tablet, see figure 12.

Aa 4: 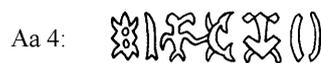

Figure 12.

Aa 4: **7-7-5 44-8 49 57** *Tuu Pipiri atua: Tava-Matua (ariki) mau Tara*. (It was) the star Canopus: king *Hotu-Matua*, (son of king) *Tara*.

Consider the record on the Tahua tablet about the arrivals of the explorers, king *Hotu-Matua* and king *Tara tahi (Tuu-ko-Iho)* on Easter Island, see figure 13.

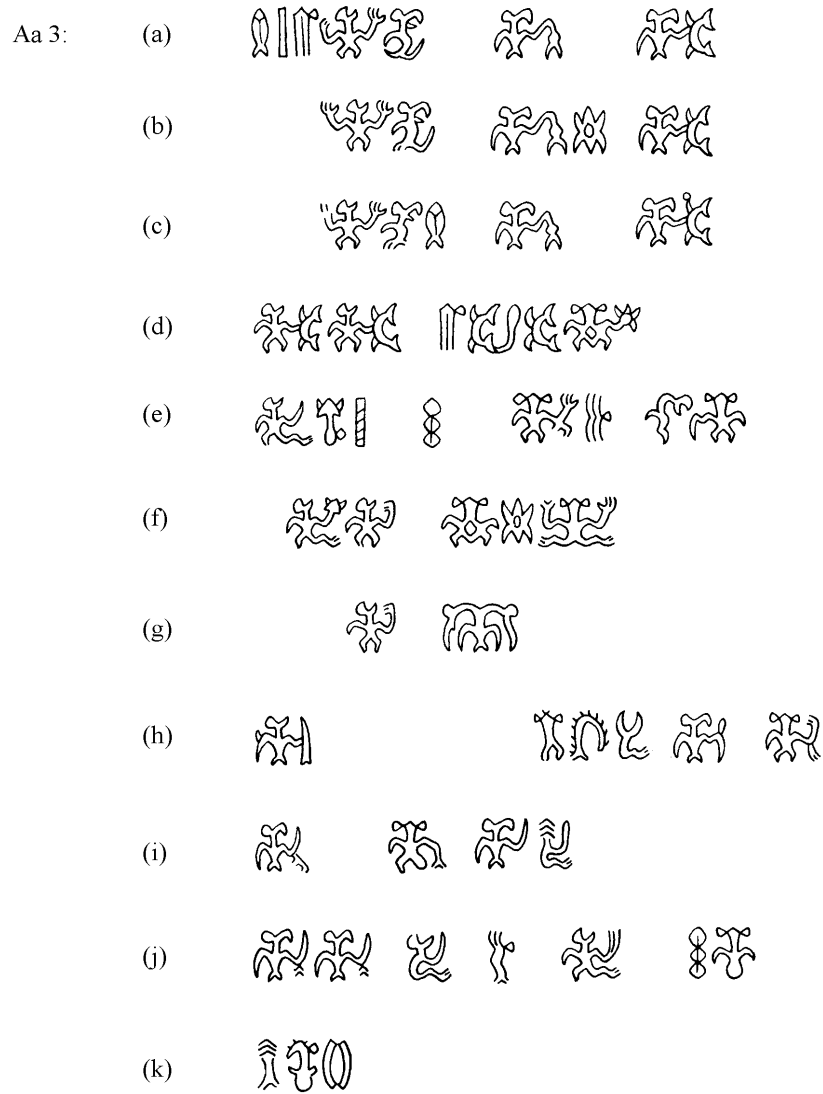

Figure 13.

Aa 3: (a) **12 4-26 6-19 44-54 44-8** *Ika Tuma, aku. Takai Tava-Matua*. (The place) of the Fish Base (the god *Tangaroa*), (the area) of the *aku* fish (near the royal residence Anakena). *Tava-Matua* voyaged (in the month *Tangaroa Uri*).
(b) **6-19 44-54 7 44-8** *Aku. Takai, tuu Tava-Matua*. (The area) of the *aku* fish (near the royal residence Anakena). *Tava-Matua* voyaged, arrived.
(c) **6-19 12 44-54 44-8 139** *Aku IKA. Takai Tava-Matua raa*. (The area) of the *aku* fish (near the royal residence Anakena). *Tava-Matua* (the solar symbolism) voyaged.
(d) **6 8 6 8 26 8 25 8 6-7** *A Matua, a Matua, maa Matua-hua, Matua-Hotu*. (It was) the great *Matua* (= *a Matua, a Matua*) (and) the son (*maa*) (called) *Matua-hua* (*Matua*-the son), *Matua-Hotu* (*Hotu-Matua* = The Creator-Father).
(e) **6-21 4 17 6-15 33 62 6** *Haka-ati te Ora-NGARU toa TANGATA*. (The ship) *Ora-Ngaru* [Swimming in August-September (*Ora, Hora*) literally] with warriors (chickens = scouts) swam.
(f) **6-21 6 6-7 69** *Haka-hoa Hotu-Moko*. *Hotu-Matua* of the tribe Moko (Hanau Momoko) put (the ship) into the water.
(g) **6 44** *Hoa Taha*. The (same) people Taha (or Tava) put (the ship) into the water.
(h) **44-5 75 14 2 44-30 6-15** *Ta-ati ko Haua Hina, Tane Hora*. In the month *Hora* (the deity) *Haua Hina* (and) *Tane* swam. (It could be the hint at figurines of those deities.)
(i) **44-5 6 44-33 2** *Ta-ati a tau Hina*. (They) swam at that lunar (*hina*) season (*tau*).
(j) **44-5-44-5 6 58 6-32 17 68** *Ta-ati-ta-ati a tahi hau (A)tea honui*. The first king, the great *Atea*, swam.
(k) **59-33 49 57** *Kau (ariki) mau Tara*. (Thus,) king *Tara* swam up.

The expression *Ika Tuma* (**12 4-26**) denotes the god *Tangaroa* (Rjabchikov 1997). The fish *aku* is included in the name of a fishing ground near Anakena (Barthel 1978: 256). The name of the fish *aku* (glyphs **6-19**) is confirmed by the determinative 'Fish' (glyph **12** *ika*). Old Rapanui *takai* 'to voyage' (glyphs **44-54**) corresponds to Marquesan *takai* 'ditto.' Rapanui *tuu* (glyph **7**) means 'to arrive.' Old Rapanui *maa* means 'son' (glyph **26**), cf. Rapanui *maanga* 'child (figuratively); chicken' < *maa-nga* as well as *maahu* 'older brother, children of an older brother' < *\*maa hua*, cf. Rapanui *hua* 'son.' Rapanui *ati* (glyphs **4** and **5**) means 'to move quickly,' cf. Rapanui *aati* 'to take a part in a race' and Maori *whati* 'to flee; to escape; to run away.' The causative prefixes *haka-* (glyphs **6-21**) and *ta-* (*taa-*) **44** are written in segments (e), (f), (h), (i), and (j).

Barthel (1978: 68) has decoded the vast rock drawing where the crew of "cocks" in the *Ora Ngaru* canoe is represented: they were the explorers floating to Easter Island. I think that these "chickens" were the designations of young warriors. Old Rapanui *toa* and *mata toa* mean 'warrior; brave man,' cf. Maori *toa* 'warrior; brave man,' Rapanui *mata toa* 'warrior,' Rapanui *moa toa* 'cock.' Furthermore, Rapanui *toa* and *mata toa* mean 'killer' (Estella 1921: 84).

It is well known that the explorers from the legendary homeland Hiva reached Easter Island in the month *Maro* (June chiefly) (Barthel 1978: 68, 73, table 4). King *Hotu-Matua* and his people deserted Hiva in the month *Hora-nui* (September chiefly) and reached Easter Island in the month *Tangaroa Uri* (October chiefly) (Ibid., p. 73, table 4; 103). In Manuscript E *Hotu-Matua* is divided into two persons, *Matua* and *Hotu* = *Hotu a Matua* [*Hotu-Matua*] (cf. Ibid., p. 98). It is a hint at two voyages of the inhabitants of Mangareva (they had the Marquesan origin) to Easter Island: *Tara tahi* (the First *Tara*) at first swam, and then *Anua-Motua* (*Hotu-Matua*) did. According to that document, the name of king *Matua* is introduced by the particle of personal names: *a Matua* (Ibid., p. 330).

In the decoded record *Hotu-Matua* is called *Tava-Matua* (glyphs **44-8**), *Matua-Hua* (glyphs **8 25**), *Matua-Hotu* (glyphs **8 6-7**) and *Hotu* (glyphs **6-7**). In this inscription king *Tara tahi*, father of *Hotu-Matua*, is called *a Matua-a Matua* (the great *Matua*: glyphs **6 8 6 8**), *Atea* (glyph **17**) and *Tara* (glyph **57**). According to Métraux (1940: between 90-91, table 2; 320-322), *Riri-Katea* was a legendary king. This name can be read as *Riri ka Atea* or *Ririka Atea* ('The Anger of *Atea* [The Whiteness]' or 'The Awaking of *Atea*,' cf. Mangarevan *rika* 'to awake and get up suddenly;' in the second case it might be an epithet of the legendary personage *Hau Maka* 'King *Maka* [The Stone]). It is well known that kings *Hotu-Matua* and *Tuu-ko-Iho* arrived in a double canoe (Ibid., p. 74).

In the memory of the Easter Islanders two different voyages of *Tara tahi* and *Hotu-Matua* turned into one joint voyage of *Hotu-Matua* and *Tuu-ko-Iho* (*Tuu-ko-Ihu*) (cf. Mangarevan *tarai* 'to carve' (< *\*tara-i*), Rapanui *tarai* 'ditto,' Maori *kātara* 'sharp point; prick' (< *\*ka tara*), and Rapanui *ihoiho* 'very hard stone').

Let us examine some segments of the previous figure, see figure 14.

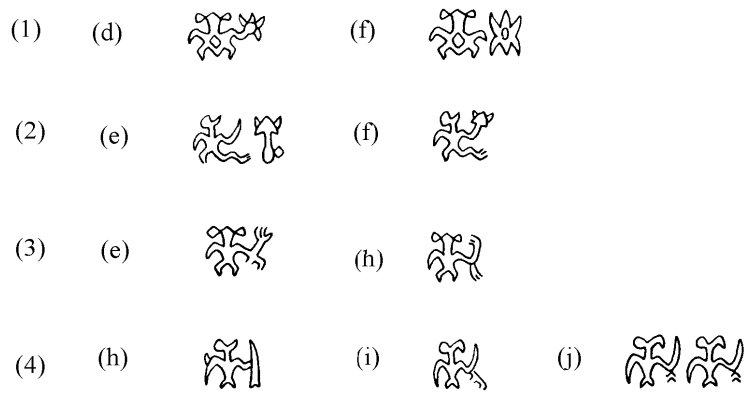

Figure 14.

One can put forward the rules of the readings of such combinations of glyphs. The first row contains glyphs **6-7** (*Hotu*) of the name of king *Hotu-Matua*: it is written as a ligature and as a normal text.

By the way, the names *Tiki-te-Hatu* and *Hotu-Matua* contain, according to Fedorova (1978: 342), the common word *hatu* = *hotu* (to bear fruits; bloom). I agree with her hypothesis. The second row contains glyphs **6-21** (the causative prefix *haka-*): it is written as a normal text and as a ligature. The third row contains glyphs **6-15** (*Hora*) as ligatures in both cases (**6- 15a**, **6-15b**). In order to show that glyph **15** depicting an arm is not included into glyph **6** depicting a human being different special signs that do not read are added to the elbow in both cases. The fourth row contains glyphs **44-5** (*taa-ati*) written as a normal text and as ligatures. Glyphs **44** and **5** in cases (i) and (j) are separated from each other with the aid of special small signs that do not read.

Amongst the explorers a personage with the odd name *Tavatava a Hua Tava* (Métraux 1940: 58-60; Barthel 1978: 65) is mentioned. I think it is the collective name of the kin of *Hotu-Matua* and his father *Tara tahi*. The name *Tavatava a Hua Tava* signifies '*Tava*, son(s) of *Tava*.' So, *Tava* (*Taha*) was the name of the tribe (cf. the name *Nga Tavake* 'Many Tropic Birds').

Consider three parallel records on the Great St. Petersburg (P), Great Santiago (H) and Small St. Petersburg (Q) tablets (cf. Rjabchikov 2014c: 4-5, figure 2), see figure 15.

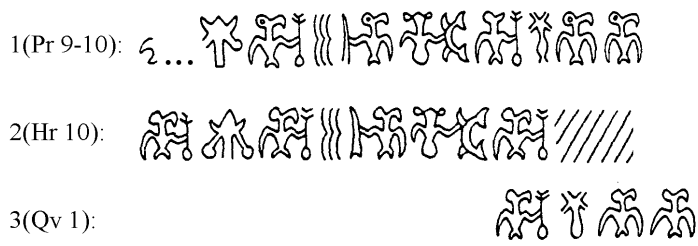

**THE RECONSTRUCTED RECORD:**

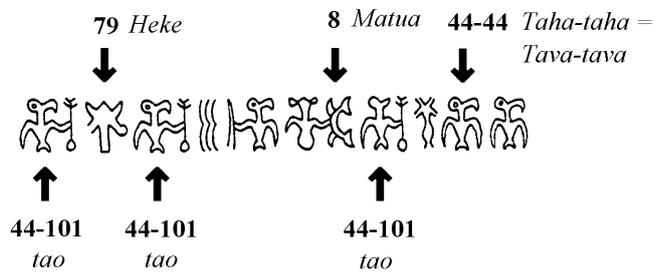

Figure 15.

The reconstructed record reads **44-101 79 44-101 33 5-44 68 8 44-101 45 44-44** *Tao Heke, tao ua Tita(h)a Honu* [*ANUA*] *Matua, tao, pu Tava-tava* [*Taha-taha*]. (It is) the forefather *Heke* (the legendary king *Tuu-Ma-Heke*), (it is) the forefather *Matua-Anua* of the dwelling (located) at the Boundary (= Anakena), (it is) the forefather-tribe (cf. Maori *pu* 'tribe; root; origin; ruler; king') *Tavatava* (*Tahataha*).

**Some Additional Remarks on the *Pipiri* Star**

The expression *tau avanga* associated with the star *Pipiri* (see above) is encoded in the record on the Santiago staff, see figure 16.

I 6: 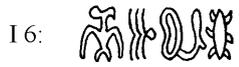

Figure 16.

I 6: **44-33 47 (102) 28** *tau avanga* the season of the grave = the rainy season

Maori *Pipiri* means 'star that is seen in the mornings a little earlier than the Pleiades.' The Maori forgot the precise correspondence of the star *Pipiri* in the sky (Best 1922: 49). It is safe to assume that this name (the bright [double] star) was a metaphor for the star *Atu tahi* (Canopus) seen unclear during the rainy season.

For calculations, I have chosen two years: 1300 A.D. and 1600 A.D. The dates of the first morning rising of Canopus before dawn were May 15, 1300 A.D. and May 24, 1600 A.D. The dates of the first morning rising of the Pleiades before dawn were May 23, 1300 A.D. and June 4, 1600 A.D.

The linkage between glyphs **9** *niva* 'darkness' and glyphs **7-7** *Tuutuu* [*Tuu Pipiri* or *Piri*] (the star *Pipiri* = Canopus) is shown in the record on the tablet Aruku-Kurenga (B), see figure 17.

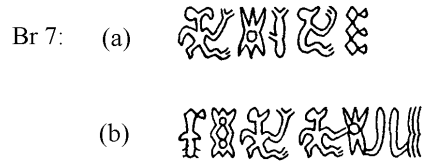

Figure 17.

Br 7: (a) **6/44 7 9 6 17** (b) **132 7-7 6 6-7 4 4-33** (a) *Hata Tuu Niva atea.* (b) *Kore Tuutuu* [*Tuu Pipiri* or *Piri*] *a Hatu atua atua/ua.* (a) The bright star from the Darkness (Grave, Winter) rose. (b) Canopus disappeared in the rays of the rising sun (the solar god *Tiki-te-Hatu*).

What is more, Aldebaran is called as a star which appeared during the winter (rainy season), too; the record is presented on the same tablet, see figure 18.

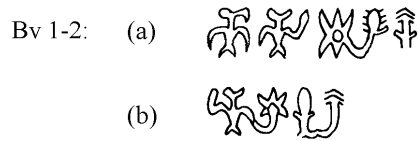

Figure 18.

Bv 1-2: (a) **44-44 7-25 9 33** (b) **44 7-25 33** (a) *Tahataha Tuu Hua niva ua.* (b) *Taha Tuu Hua ua.* (a) The star Aldebaran (appeared first in the rainy season) turned. (b) The star Aldebaran (appeared first in the rainy season) turned.

Barthel (1978: 39, 41) has compared the Rapanui place name *Te Piringa Aniva* with the name of the star *Pipiri* (cf. Tuamotuan *Piringa-o-tautu*). The last name literally means 'the star *Pipiri* of a (certain) season of its existence' (*Piringa o tau tuu*). Besides, Old Rapanui *aniva* means 'dark; black' (Ibid., p. 44) < *a niva*; cf. Maori *niwaniwa* 'dark; deep black.' So, the name *Te Piringa a Niva* means 'Canopus (appeared) from the Darkness.'

**The Search for Allographic Signs to Read Astronomical Records Correctly**

Barthel (1958: 202-203) has recognised glyphs **B2** *maitaki*, *maharonga*, *inoino*, *matariki* and **B20** *mata* (in his own classification) as different signs. In contrast to that approach, I always read them (glyphs **17** and **18** in my own classification) as *te* and *tea* (Rjabchikov 1987: 362, figure 1; 364, figure 2).

Consider the record on the Great Santiago tablet, see figure 19.

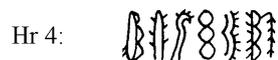

Figure 19.

Hr 4: **18 25 4 14 17 50 17 24** *Te (h)ua atua HAU, te hi, te arii.* The royal staff *ua*, the solar rays, the king.

Four titles (*atua* [lord], *hau* [king], *hi* [the sun] and *arii* [king]) and one material symbol (*ua*) of the royal authority are set out in this didactic record. The different variants of glyph **17** (**18**) *te* 'definite article' are taken down. Old Rapanui *hi* 'solar rays' correlates with Maori *hihi* 'ray of the sun' and *hi* 'dawn.'

During Cook's expedition on Easter Island in 1774 A.D. the terms *hariikii* [*hariki* or *'ariki*] and *arii* 'chief' [*hareekee*, *aree*] were put down (Forster 1777: 588; Schuhmacher 1978: 4). I suggest that two versions of this term co-existed because of the alternations of the sounds *k/-*. A chief wearing a hat decorated with black feathers of frigate birds told the Europeans that he was called *Ko Tohi Tai* [*Ko Toheetai*] (Forster 1777: 589). I think that it was king *Kai Makoi* the First wearing the royal hat (*hau teketeke*). His high title, *Ko to hi*, means 'The Addition of the Solar Rays,' and his name was *Tai*, otherwise *Kai* because of the alternations of the sounds *t/k* (cf. also Rapanui *tai* 'song' and *kaikai* 'string figure').

Consider two pair of parallel records, the first pair on the Small Santiago (G) and Keiti (E) tablets, and the second pair on the Small and Great St. Petersburg tablets, see figure 20. Again, different versions of glyph **17** (**18**) are recognised.

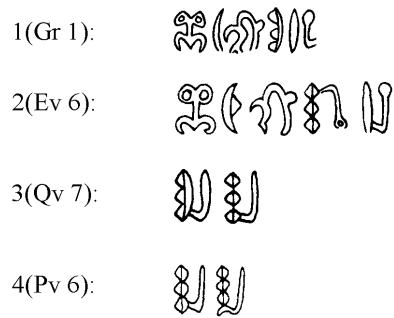

Figure 20.

1 (Gr 1): **60-23 3 44b-18 30-4** *Mata Ura, hina, Tua Tea nati.* The conjunction of Mars (The Red Eye), the moon and Venus (The Turn from the Light/Vision; cf. Tongan *Tapu Kitea*) occurred.
2 (Ev6): **60-23 3 44b-17 65 30-4** *Mata ura, hina, Tua tea RANGI nati.* Ditto.
3 (Qv 7): **18-4-17-4** *tetutetu* big
4 (Pv 6): **17-4-17-4** *tetutetu* ditto

Consider again the records about Canopus that contain different variants of glyph **17** (**18**), see figure 21. Here figures 1, 3 and 6 are used.

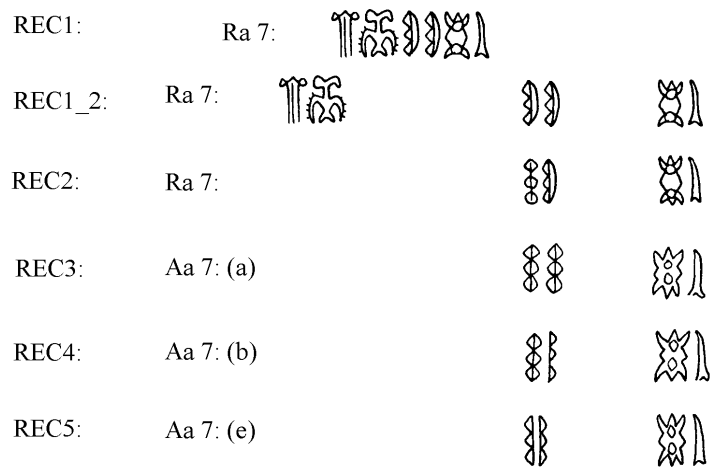

Figure 21.

The record about Canopus: **26-44** *Mata* **17var-17var** *teatea* (= *ka*) **7-7** (*tuu*) *Pipiri* **5** *Atua*, and some parallels.

Rapanui *piri* means 'to join,' and Maori *pipiri* 'to cling together.' It is interestingly to note that in the "*Apai*" text the words about Canopus are introduced by the word *taku*, cf. Rapanui *taku* 'to predict' and *tataku* 'to count; to add.' Mangarevan *Pipiri* signifies 'the season about June; the cold season; the winter, the season when all things rest.' The word *anua* in the Mangarevan name *Anua-Motua* and Rapanui name *Matua-Anua* of king *Hotu-Matua* means 'coldness; cold,' cf. Mangarevan *anu* 'ditto.'

Consider the record on the Small St. Petersburg tablet, see figure 22.

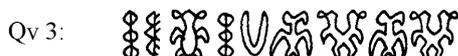

Figure 22.

Qv 3: **… 17-17 68 17 61 44 6 44 6** … *teatea Honu; te hina taha ha, taha ha.* … the Pleiades appeared; the moon passed during four nights (and next) four nights (= it was the first quarter of the moon).

Again, two versions of glyph **17** are presented. The second glyph **17** is a transitive form from glyph **17** (a half of the sign) to glyph **18**.

**How I Have Suddenly Realised Two *Rongorongo* Texts**

Consider the record on the Keiti tablet, see figure 23. Now this text is understandable.

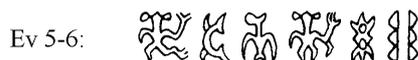

Figure 23.

Ev 5-6: **6 8 44 6-15 7-7 17-17** *A Matua Tava Hora Tuu Pipiri* (= *Tuutuu*) *teatea*. (King) *Matua* [= *Hotu-Matua*, *Matua-Anua*, *Anua-Motua*] (from the tribe) *Tava* (= *Tavake*) (was connected with) the season *Hora* (*Hora-iti*, *Hora-nui*, August and September chiefly) (and) glistening Canopus.

According to this record in a lesson book from the *rongorongo* school of king *Kai Makoi* the First, the voyage of king *Hotu-Matua* from Hiva to Easter Island started in the season *Hora* (August-September chiefly) and was connected with Canopus.

Consider the record on the Great Santiago tablet, see figure 24. It has become clear as well.

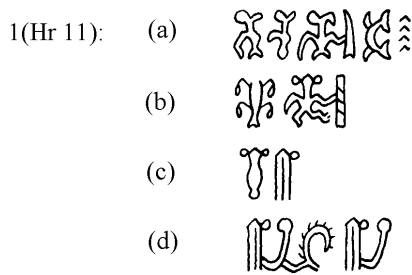

Figure 24.

Hr 11: (a) **19-19 44 5 8 24** *Kuukuu Tava atua Matua arii*, (They were) *Kuukuu*, (a son) of *Tava*-lord-*Matua*-king,
(b) **25 6-4** *hua Hotu.* (and other) sons of *Hotu*.
(c) **56 26** *Poa maa* The children (*maa*)
(d) **26-4 14 26-4** *Matua hau Matua.* of *Matua*-king-*Matua* landed (*poa*).

The first five glyphs of this text have earlier been deciphered (Rjabchikov 2012b: 28, figure 5). The high-ranking name *Kuukuu* of the older son of *Tava* (*Hotu-Matua*) is associated with the power symbol-ism, cf. Rapanui *kuukuu* 'to call chickens (of a hen).' I have demonstrated that the word *matua* could be written as an ideogram (glyph **8**) and as quasi-syllables (glyphs **26-4**) (Rjabchikov 2014b: 170, figure 6, fragments 3 and 4). One can translate the term *poa* as 'to land,' cf. Rapanui *poa* 'to tie a boat.'

Let us consider the rock drawing at Complex A, Orongo (Lee 1992: 104, figure 4.101, (1)). Here a double canoe is depicted. That symbol is incised together with glyph **149** (= the inverted variant) *Hotu* or *Hatu*. Hence, it is a report about the arrival of the double canoe of *Hotu-Matua* on Easter Island.

In conformity with Manuscript E (Barthel 1978: 56, 73, table 4), it is easily to calculate, the explorer *Makoi* came to a spot named *Te Piringa a Niva* in the month *Maro* (June chiefly), the rainy season, on the 16th day after the explorers' arrival on Easter Island. One can anticipate therefore that that place was the small observatory to watch Canopus with special emphasis on its heliacal rising. On the other hand, one can advance that the explorers (king *Tara tahi* with his crew in actuality) saw Canopus at first in the morning somewhere in the ocean, and in 16 days they reached the coast of Easter Island. Of course, they could arrive on the island soon after the heliacal rising of Canopus.

Let us consider the rock drawing of the ship *Ora Ngaru* with the explorers (Lee 1992: 98, figure 4.91). In compliance with Barthel (1978: 68), the bow of that canoe on that picture was decorated with the head of the bird. I have read this symbol as glyph **44** *Tava* (*Taha*). On the keel of the canoe one can reckon 16 cupules, cf. the information about 16 days calculated above. On the board the canoe there are symbols of six chickens and glyph **39** *raa* (the sun).

The structure of the five explorers' names, *Kuukuu*, *Ringiringi*, *Nonoma*, *Uure*, and *Makoi*, the childen of *Tava* (= *Hotu-Matua*), has been analysed above. Besides, among the explorers were two children of *Hau Maka*, namely *Ira* as the first-born son and *Raparenga* (Ibid., p. 54). The spirit of *Hau Maka* dwelt at Mount Rano Kau on Easter Island (Fuentes 1960: 696), therefore it could be a king indeed who arrived on the island from Hiva.

The name of *Ira* is written with the help of glyph **39** *raa* (the sun). This name sounded as *(H)i Ra* 'Rays of the Sun.' Near *Ira* the "chicken" *Rapa Renga* is depicted. The name of the latter hero signifies 'The Beautiful (Ceremonial) Paddle *Rapa*.' Judging from the position of the latter personage, he steered the canoe. Both explorers as the children of *Hau Maka* (= *Tara tahi*, *Tuu-ko-Iho*) are shown in the upper part of the picture.

The children of *Tava* (*Hotu-Matua*) are revealed in the picture at the bottom. The biggest "chicken" is *Kuukuu* gathering other "chickens." The smaller companions are combined in two lines.

The "chicken" in the top row on the left is *Uure* 'The Son.' The "chicken" on the right is *Makoi* (cf. Rapanui *makoi* 'certain plant; its fruit; nut: the late term').

The "chicken" in the second row on the left is *Nonoma*, and the cursive variant of glyph **26** *maa* is seen in its tail, cf. Rapanui *nonoma* 'very shiny' and *maa* 'bright; clean; clear.' The "chicken" on the right is *Ringiringi*, and his wings resemble glyph **120** *ngii*. In this case, the name *Ringi* (*Ringiringi*) reads *Ri* (= *Riri*) *(A)ngi* 'The Anger of the Bright Sun.' At that rate we can decode the name *Meriri* of the people of king *Tara tahi*. First of all, this ethnicon is comparable with the name of the legendary king *Riri-Katea* known in the Rapanui mythology. Besides, it reads *Mea Riri* 'The Red Colour in Anger,' and that colour could be an attribute of the gods *Tangaroa* and *Tane* (*Tiki*).

## Conclusions

Watchings of Canopus as a herald of the winter were important duties of ancient priests-astronomers on Easter Island. All the analysed data witness that this star was observed during the first and second voyages from Mangareva to the island. The names of king *Hotu-Matua* (*Anua-Motua*) and his father *Tara tahi* have been decoded. Several *rongorongo* records from the Esteban Atan manuscript have been deciphered advantageously. The new view at a painted barkcloth figurine connected with the bird-man cult has been offered. Some data about watchings of Aldebaran, the Pleiades, the sun, the moon, Venus and Mars have been collected as well.

Mulloy, W., 1973. Preliminary Report of the Restoration of Ahu Huri a Urenga and Two Unnamed Ahu of Hanga Kio'e, Easter Island. *Bulletin 3*, Easter Island Committee. New York: International Fund for Monuments.

Mulloy, W., 1975. A Solstice Oriented *Ahu* on Easter Island. *Archaeology and Physical Anthropology in Oceania*, 10, pp. 1-39.

Orliac, C. and M. Orliac, 1995. *Bois sculptés de l'île de Pâques*. Marseilles: Editions Parenthèses, Editions Louise Leiris.

Popova, T., 2015. The Rapanui *rongorongo* Schools. Some Additional Notes. *Anthropos*, 110(2), pp. 553-555.

Rjabchikov, S.V., 1987. Progress Report on the Decipherment of the Easter Island Writing System. *Journal of the Polynesian Society*, 96(3), pp. 361-367.

Rjabchikov, S.V., 1997. A Key to the Easter Island (Rapa Nui) Petroglyphs. *Journal de la Société des Océanistes*, 104(1), p. 111.

Rjabchikov, S.V., 2000. La trompette du dieu Hiro. *Journal de la Société des Océanistes*, 110(1), pp.115-116.

Rjabchikov, S.V., 2001. *Rongorongo* Glyphs Clarify Easter Island Rock Drawings. *Journal de la Société des Océanistes*, 113(2), pp. 215-220.

Rjabchikov, S.V., 2010a. Rapanuyskie zhretsy-astronomy v Tongariki. *Visnik Mizhnarodnogo doslidnogo tsentru "Lyudina: mova, kul'tura, piznannya",* 24(1), pp. 69-76.

Rjabchikov, Sergei V., 2010b. On the Methodology of Decoding the *Rongorongo* Script: Statistical Analysis or Distributive One? *Polynesian Research*, 1(3), pp. 3-35.

Rjabchikov, S.V., 2010c. Ocherk po istorii ostrova Paskhi. *Visnik Mizhnarodnogo doslidnogo tsentru "Lyudina: mova, kul'tura, piznannya",* 27(4), pp. 35-53.

Rjabchikov, S.V., 2011. Canopus and the Pleiades in Records on the Tahua Tablet. *Polynesian Research*, 2(1), pp. 11-12.

Rjabchikov, S.V., 2012a. The rongorongo Schools on Easter Island. *Anthropos*, 107(2), pp. 564-570.

Rjabchikov, Sergei V., 2012b. An Outline of Polynesian Language Development. *Polynesian Research*, 3(4), pp. 11-31.

Rjabchikov, S.V., 2013. The Astronomical and Ethnological Components of the Cult of Bird-Man on Easter Island. arXiv:1309.6056 [physics.hist-ph].

Rjabchikov, S.V., 2014a. *On the Observations of the Sun in Polynesia*. arXiv:1407.5957 [physics.hist-ph].

Rjabchikov, S.V., 2014b. The God Tinirau in the Polynesian Art. *Anthropos*, 109(1), pp. 161-176.

Rjabchikov, S.V., 2014c. The Sea Route from the Marquesas to Mangareva and Then to Easter Island: New Data. *Polynesia Newsletter*, 3, pp. 2-14.

Rjabchikov, S.V., 2014d. The Sea Route from the Society Islands to Easter Island. Early Mangarevan-Marquesan Records on Easter Island: New Data. *Polynesia Newsletter*, 4, pp. 2-12.

Rjabchikov, S.V., 2015. *Easter Island: the Tongariki and Mataveri Solar Observatories Used a Common Methodology*. arXiv:1508.07100 [physics.hist-ph].

Rjabchikov, S.V., 2016a. *The Ancient Astronomy of Easter Island: Venus and Aldebaran*. arXiv:1604.03037 [physics.hist-ph].

Rjabchikov, S.V., 2016b. *The Ancient Astronomy of Easter Island: Aldebaran and the Pleiades*. arXiv:1610.08966 [physics.hist-ph].

Routledge, K., 1998. *The Mystery of Easter Island*. Kempton: Adventures Unlimited Press.

Schuhmacher, W.W., 1978. Un vocabulario inédito de la Isla de Pascua: compuesto por Johann Reinhold Forster durante el segundo viaje de Cook. *Moana, Estudios de Antropología Oceanica*, 1(12), pp. 1-14.

Te Rangi Hiroa (Buck, P.), 1949. *The Coming of the Maori*. Wellington: Maori Purposes Fund Board.

Thomson, W.J., 1891. Te Pito te Henua, or Easter Island. Report of the United States National Museum for the Year Ending June 30, 1889. *Annual Reports of the Smithsonian Institution for 1889*. Washington: Smithsonian Institution, pp. 447-552.
17